# Bistable Amphoteric Native Defect Model of Perovskite Photovoltaics

W. Walukiewicz[1,2], I. Rey-Stolle[2,3], G. Han[4], M. Jaquez[2,5], D. Broberg[1], W. Xie[1], M Sherburne[1], N. Mathews[4,6] and M. D. Asta[1,2]

[1] Department of Materials Science and Engineering, University of California, Berkeley, California 94720, USA

[2] Materials Sciences Division, Lawrence Berkeley National Laboratory, Berkeley, CA, 94720, USA

[3] Solar Energy Institute, Technical University of Madrid (UPM), Madrid, Spain

[4] Energy Research Institute @NTU (ERI@N), Nanyang Technological University, Research Techno Plaza, X-Frontier Block, Level 5, 50 Nanyang Drive, 637553, Singapore

[5] Department of Mechanical Engineering, University of California, Berkeley, California 94720, USA

[6] School of Materials Science and Engineering, Nanyang Technological University, Nanyang Avenue, 639798, Singapore

**Abstract**

**The past few years have witnessed unprecedented rapid improvement of the performance of a new class of photovoltaics based on halide perovskites. This progress has been achieved even though there is no generally accepted mechanism of the operation of these solar cells. Here we present a model based on bistable amphoteric native defects that accounts for all key characteristics of these photovoltaics and explains many idiosyncratic properties of halide perovskites. We show that a transformation between donor-like and acceptor-like configurations leads to a resonant interaction between amphoteric defects and free charge carriers. This interaction, combined with the charge transfer from the perovskite to the electron and hole transporting layers results in the formation of a dynamic *n-i-p* junction whose photovoltaic parameters are determined by the perovskite absorber. The model provides a unified explanation for the outstanding properties of the perovskite photovoltaics, including hysteresis of *J-V* characteristics and ultraviolet light-induced degradation.**

Solar cells based on Hybrid Organic-Inorganic Perovskites (HOIP) were first reported less than ten years ago. Although initially a solar power conversion efficiency of 3.9%[1] was reported, subsequent research led to a several-fold improvement[2,3]. These new developments generated enormous interest and initiated an unprecedented rise in research on HOIPs, both for photovoltaic and light emitting device applications. For solar cell applications it has led to record power conversion efficiency of more than 22%[4]. These spectacular improvements resulted from the painstaking experimentation of numerous research groups worldwide. Despite these rapid experimental developments, there is still no commonly accepted physical model that accounts for the outstanding properties of the HOIPs and explains the mechanism of perovskite solar cell operation. The progress in the understanding of various aspects of perovskite materials and perovskite-based PVs has been, at different stages, reviewed thoroughly in several publications[5-10].

It is now commonly recognized that the extraordinary properties of halide perovskites as photovoltaic materials result from their unique defect properties. First-principles calculations have been performed to describe the nature of native point defects in HOIPs, leading to varying descriptions of the defect physics[11-13]. However, a commonly accepted experimental fact remains that the synthesized HOIP materials lack non-radiative trap states that can result from certain types of point defects. A widely used explanation for this property is the concept of "defect tolerance" wherein the dangling bond states caused by simple vacancies produce charge transition states which are very close to the band edges[14,15]. In the case of MAPbI$_3$, the iodine vacancy donor ($V_I$) has a (+/0) charge transition state close to the conduction band edge (CBE) and the lead vacancy ($V_{Pb}$) and MA vacancy ($V_{MA}$) acceptors have their (0/-) charge transition states close to the valence band edge (VBE). This unique feature of perovskites suggests that any dangling bond-like defect or associated defect complexes will have transition states close to the band edges.

In this paper, we propose a model of the operation of HOIP photovoltaic devices that provides an explanation for the various operational phenomena noted in them. It is based on the concept of amphoteric native defects, i.e., defects whose formation energy as well as their nature (donor or acceptor) depend on the location of the Fermi energy. Originally the concept of amphoteric defects was introduced to explain effects of native (or intrinsic) defects on the properties of semiconductor materials including the pinning of the Fermi energy on semiconductor surfaces, formation of Schottky barriers and limitations on the effective doping and stabilization of the Fermi energy in heavily defective semiconductors[16-18]. More recently the concept of amphoteric native defects has been used to devise methods to control defect incorporation in compound semiconductors[19].

The key feature of amphoteric defects is the concept of the Fermi level stabilization energy ($E_{FS}$) that is universally located at about 4.9 eV below the vacuum level for all semiconductors[20]. The origin of the constant value of the $E_{FS}$ on the absolute energy scale has been attributed to the material independent location of the charge transition states of highly localized dangling bond like defects[20,21]. Most typically the dangling bonds are represented by simple vacancies although it has been shown that they can form by hydrogen interstitials attaching themselves to either anion or cation sites of compound semiconductors[22].

**Bistable Amphoteric Native Defects**

In the following, we consider a simple amphoteric defect system with the (+/0) donor charge transition state close to the CBE and (0/-) acceptor charge transition state close to the VBE as in CH$_3$NH$_3$PbI$_3$. The schematic formation energy diagram for the defect is shown in Fig. 1 (a). It should be emphasized that the donor and acceptor in Fig. 1 (a) are two different stable configurations of the same defect. The transition between donor and acceptor-like configuration is associated with the defect site relaxation. The process is illustrated in Fig. 1(b) that shows a configuration diagram of the energy of an amphoteric defect in the neutral state (black line). Such a diagram is a standard way of representing the energy of a defect undergoing structural reconstruction. The configuration coordinate represents the combined movement of atoms participating in a local lattice reconstruction. This could be a breathing mode of all atoms surrounding the defect, a symmetry breaking movement of the nearest neighbors[20,21] or movement of H-interstitial from forming a bond with the cation or anion atom[22]. Here we assume that, as is shown in Fig. 1(b), the energy of the neutral state of the amphoteric defect has two minima one corresponding to the acceptor-like ($R_A$) and the other to the donor-like ($R_D$) configuration. The energy barrier, $E_B$, between the minima determines the minimum temperature for reaching the equilibrium distribution between the neutral defects in donor-like and acceptor-like configurations.

At this point, it is important to highlight the unique features of the considered physical system. As is shown in Fig. 1 (c), it consists of two interacting subsystems; a semiconductor with a direct band gap, $E_g$ and the system of bistable amphoteric native defects (BANDs). In the ground state, the semiconductor is in a neutral or insulating state with no electrons in the conduction band and no holes in the valence band. The excited state is produced when an electron is transferred from the valence to the conduction band producing an electron ($e_{PVSK}$) and hole ($h_{PVSK}$) pair. This is the case of a positive energy band gap. The BAND subsystem shows the opposite behavior; it consists of electronic states that are positively or negatively charged when the subsystem is in the ground state. An excitation transfers the defects into the neutral state, which is the excited state of the BAND subsystem. Thus, it can be argued that the BANDs behave as a subsystem with a "negative" gap. This unusual property is a direct consequence of the negative U behavior of the BANDs in the acceptor-like configuration[21]. The two subsystems can exchange energy facilitating migration of an excitation. A critically important feature of those two interacting subsystems is that the BANDs serve as a channel to transfer electrons (holes) in the conduction (valence) band into holes (electrons) in the valence (conduction) band in the following reversible process,

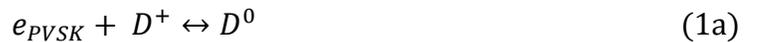
$$e_{PVSK} + D^+ \leftrightarrow D^0 \tag{1a}$$

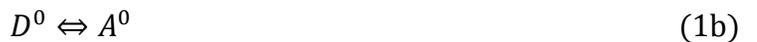
$$D^0 \Leftrightarrow A^0 \tag{1b}$$

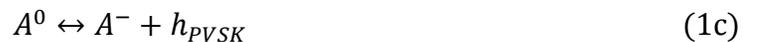
$$A^0 \leftrightarrow A^- + h_{PVSK} \tag{1c}$$

In this set of reactions, schematically illustrated in Fig. 1(c) an electron (hole) in the conduction (valence) band has been replaced by a hole (electron) in the valence (conduction) band and one of the positively charged BAND donors (acceptors) has been transformed into a negatively (positively) charged acceptor (donor). Note that apart from the reaction (1b) the other two reactions involve trapping and de-trapping of free charges by defects in the donor and/or acceptor configuration. The reaction (1b) represents (shown in Fig. 1(b)) transformation of a neutral BAND between two stable defect configurations. The key consequence of this chain of reactions is that it creates a balance between electron and hole concentration. As will be shown in a forthcoming paper this has profound consequences for light emission and balanced charge transport in perovskites.

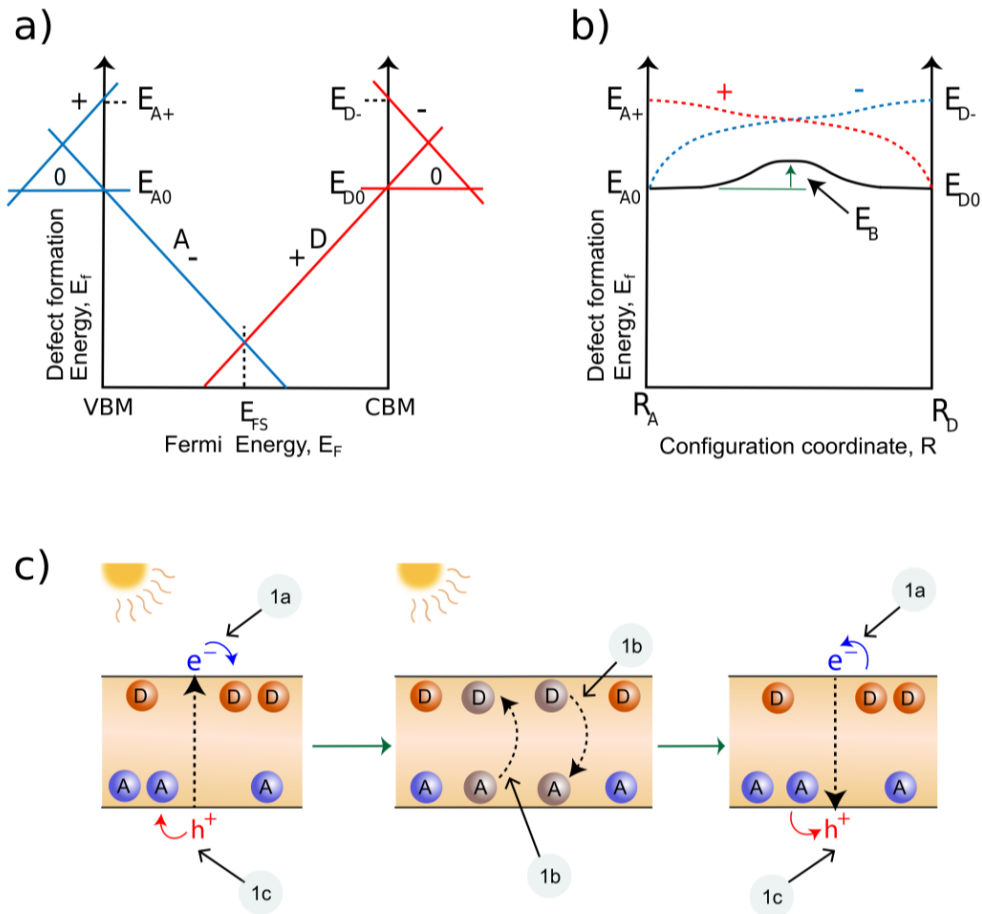

**Figure 1** (**a**) Fermi energy dependence of the amphoteric defect formation energy in different charge states. (**b**) Configurational diagram of the defect formation energy. A neutral defect can transform between donor-like ($R_D$) and acceptor-like ($R_A$) configuration by surmounting the barrier, $E_B$. (**c**) Resonant energy transfer from photoexcited electron (e⁻) hole (h⁺) pair (left panel) into positively charged donor (D), negatively charged acceptor (A) resulting in formation of a neutral defect in donor (D) (reaction 1a) and acceptor (A) (reaction 1c) configuration. The balance between donor-like and acceptor-like defect configurations is maintained by reaction 1b (middle panel). After switching off the light, the energy stored by neutral defects is transferred back to free electrons and holes that recombine restoring the system to the ground state with equal concentrations of positively charged donors and negatively charged acceptors (right panel).

To infer the effects of free charge carriers on the BANDs, we note that, as is shown in Fig. 1(a), the formation energy and the defect concentrations in donor and acceptor configurations depend on the Fermi energy and thus also on the charge carrier concentrations. The derivation shown in the Supplemental Information (SI) section yields the following expressions, for the concentration of the BANDs in the donor, $N_D$ and acceptor, $N_A$ configuration,

$$N_D = \frac{N_d A N_c p^2}{A N_c p^2 + N_v n_i^2} \tag{2}$$

$$N_A = \frac{N_d N_v n^2}{N_v n^2 + A N_c n_i^2} \tag{3}$$

Where $N_d$ is the BANDs concentration, $N_c(N_v)$ is the density of states in the conduction (valence) band, $n_i^2 = N_c N_v exp\left(-\frac{E_g}{kT}\right)$, $E_g$ is the band gap of the material, $n$ ($p$) is the electron (hole) concentration in the conduction (valence) band and,

$$A = exp\left(\frac{E_{A0} - E_{D0}}{kT}\right) \tag{4}$$

Equations (2) and (3) are important for the understanding of many properties of halide perovskite materials, as they show that the concentration of BANDs in the donor (acceptor) configuration is directly related to the concentration of free holes (electrons) in the valence (conduction) band.

**Hybrid Organic-Inorganic Perovskite Solar Cell Operation**

In the preceding discussion, we have considered a bulk halide perovskite material with a specific concentration of BANDs. In the case of vacancies, the equilibrium bulk concentration of the defects depends on crystal non-stoichiometry and the Fermi energy. It can also be affected by the presence of extended defects, e.g. crystalline grain boundaries. Obviously, the concentration of defects can be significantly higher close to external surfaces and interfaces with other materials where the stoichiometry can be affected by chemical reactions and changes in the Fermi energy resulting from charge transfer[23-25].

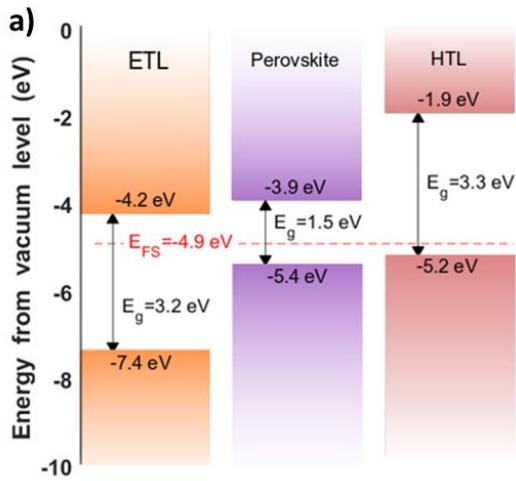

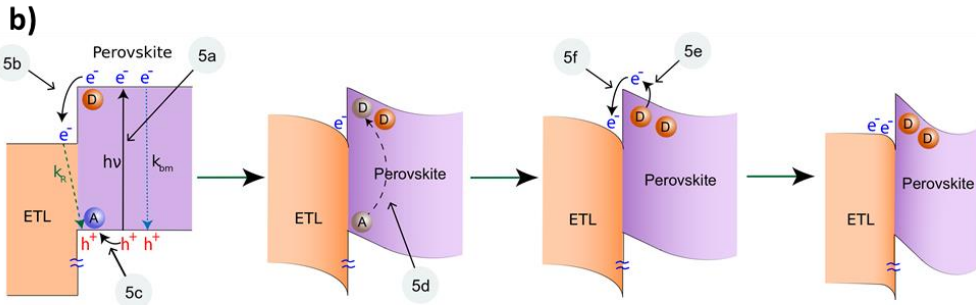

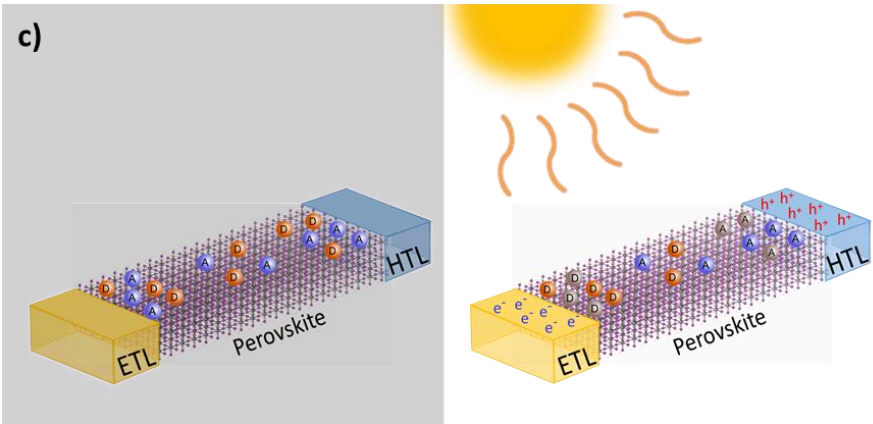

**Figure 2** (**a**) Typical PVSC structure with a TiO$_2$ ETL and Spiro:OMeTAD HTL with higher defect concentrations at the interfaces marked by the shaded areas. The Fermi level stabilization energy E$_{FS}$ is represented by red line. (**b**) Electron transfer induced defect transformation and formation of n-type perovskite region adjacent to ETL. Electrons photoexcited to the perovskite conduction band (reaction 5a) transfer to lower energy CBE of ETL (5b) whereas the photoexcited holes that accumulate at the interface with ETL neutralize negatively charged BAND acceptors (5c) creating an excess of neutral BANDs in the acceptor configuration. The neutral acceptors relax into neutral donor configuration (5d). The donors become positively charged by transferring electrons to the ETL (5e) and (5f). Consequently, all the BANDs close to the interface assume the donor configuration and transfer electrons to the ETL leading to the formation of a positively charged depletion layer in the perovskite balanced by the negative charge of electrons accumulated at the ETL/perovskite interface. Thus, the layer adjacent to the ETL is becoming *n*-type with the perovskite CBE located close the Fermi energy, schematically shown in Fig. 2(b). Analogous arguments, with electrons (holes) replaced by holes (electrons) and donors (acceptors) by acceptors (donors), apply to hole transfer and formation of the p-type layer on the HTL side. (**c**) Illustration of the dynamic n-i-p junction with n-type (p-type) region formed by the light induced electron (hole) transfer. The interfacial regions are equivalent to optically charged capacitors.

In a standard perovskite solar cell (PVSC) the absorber (MAPbI$_3$) is cladded with charge selective contact layers; an electron transport layer (ETL) on one side and a hole transport layer (HTL) on the other[26,27]. Figure 2(a) shows a typical PVSC configuration with TiO$_2$ as ETL and Spiro:OMeTAD as HTL. It should be noted that a number of materials including TiO$_2$[3], ZnO[28], SnO$_2$[29] and PCBM[30] have been shown to work as ETLs, and an even larger variety of materials including Spiro:OMeTAD[2], PTAA[31], CuSCN[32,33], CuI[34] and carbon[35] have been used as HTLs.

A striking characteristic of a PVSC is that device structures utilizing ETLs and HTLs with different locations of the CBE and the VBE energies tend to produce similar, close to or larger than 1 eV, open circuit voltages ($V_{oc}$'s), strongly indicating that the $V_{oc}$ is determined by the band gap of the perovskite absorber ($E_g$ = 1.55 eV) rather than the Fermi level difference between ETL and HTL[7]. This, however, is inconsistent with the fact that the perovskite absorber is an undoped insulator with the Fermi energy in the band gap[36]. There is no obvious way to address this conundrum. Here we show that the bistability and amphoteric behavior of BANDs allow construction of a simple model of the PVSC's operation.

As has been discussed in the preceding section, in the absence of any surface or interfacial interactions, BANDs stabilize the Fermi energy at $E_{FS}$ in the gap of the perovskite layer. However, the situation changes when the perovskite is in contact with materials that can accept either electrons from the CBE or holes from the VBE of the perovskite layer. Consider first the case of the planar MAPbI$_3$/TiO$_2$ interface shown in Fig 2(a). Typically, undoped TiO$_2$ is a slightly *n*-type semiconductor with the Fermi level close to the CBE. Also, since, as in other semiconductors, the surface of TiO$_2$ is pinned in the gap there is no significant charge transfer between TiO$_2$ and the insulating perovskite in the dark.

In contrast, under illumination free charges can transfer between perovskite and the ETL. The process is illustrated in Fig 2 (b) and represented by the set of reactions:

$$\text{Light on} \qquad h\nu \rightarrow e_{PVSK} + h_{PVSK} \tag{5a}$$

$$e_{PVSK} \rightarrow e_{ETL} \tag{5b}$$

$$h_{PVSK} + A^- \rightarrow A^0 \tag{5c}$$

$$A^0 \Rightarrow D^0 \tag{5d}$$

$$D^0 \rightarrow D^+ + e_{PVSK} \tag{5e}$$

$$e_{PVSK} \rightarrow e_{ETL} \tag{5f}$$

The key difference between the reactions 1a to 1c and reactions 5a to 5f is that the former reactions describe a reversible defect transformation as represented by reaction 1b, whereas, the latter describes an irreversible defect transformation from an acceptor to a donor configuration (5d) driven by excess holes in the valence band accumulating at the perovskite/ETL interface. The formation of the *n*-type layer in the perovskite is associated with the absorption of photons (5a) whose energy is stored by the electrons in the conduction band of the ETL (5f).

The discussed mechanism of the charge transfer induced formation of *n*-type perovskite layer is a dynamic process that depends on a variety of parameters such as band offsets, total defect concentration as well as recombination and charge transfer rates for electrons and holes Thus, in the notation of Fig. 2(b) the formation of the *n*-type perovskite interface region requires that the electron transfer rate from the perovskite to the ETL, $k_t$, and the transfer rate between donor and acceptor configurations of neutral BANDs are not much slower than the rate of bimolecular recombination $k_{bm}$ and the rate of interlayer recombination $k_R$ of electrons in the ETL with holes in the perovskite.

The process of electron transfer to the ETL slows down when the perovskite CBE approaches the Fermi energy in the ETL, and an increasingly smaller fraction of electrons is transferred to the ETL. The downward movement of the perovskite CBE towards the Fermi energy leads to a higher electron concentration and a higher concentration of neutral BANDs. Eventually, the process stops when the concentration of electrons in the perovskite CBE becomes equal to the concentration of electrons transferred to the ETL. This condition defines the location of the Fermi energy in the perovskite next to the interface with the ETL. Further away from the interface where there is no charge transfer, the perovskite recovers its bulk properties where BANDs in donor and acceptor configurations form with comparable rates resulting in a compensated material.

In order to estimate the minimum location of the CBE relative to the Fermi energy in the "*n*-type" perovskite interfacial layer we assume that in the *n*-type ETL the Fermi energy, $E_F$ is located close to the CBE, ($E_{ETL}$), and that the total concentration of BANDs in the perovskite layer adjacent to the ETL equals $N_d$. The charge transfer between the perovskite and the ETL is controlled by the energy difference $\Delta = E_c - E_F$, where $E_c$ is the perovskite CBE energy. Note that the energy difference $\Delta$ is changing with the distance from the interface and is decreasing with increasing thickness of the depletion layer. At any distance from the interface, the concentration of charge transferred from perovskite to the ETL is given by,

$$n_t = N_d[1 - \exp\left(-\frac{\Delta}{kT}\right)] \qquad (6)$$

whereas the concentration of electrons in the conduction band is equal to,

$$n = N_C \exp\left(-\frac{E_c - E_F}{kT}\right) = N_C \exp\left(-\frac{\Delta}{kT}\right) \qquad (7)$$

The charge transfer induced shift of the perovskite CBE stops when $n = n_t$ or

$$\Delta = E_c - E_F = kT ln(\frac{N_d + N_C}{N_d}) \tag{8}$$

It is seen from Eq. 8 that the minimum location of the perovskite CBE relative to $E_F$ depends on the ratio of the total concentration of BANDs, $N_d$ and the conduction band density of states, $N_C$. Thus for $N_d \gg N_C$, the layer adjacent to the ETL behaves as an *n*-type semiconductor with $E_c \approx E_F$ whereas for $N_d \ll N_C$ the value of $\Delta$ is large meaning that there is no shift of the $E_c$ towards $E_F$. These considerations strongly emphasize the importance of a large enough concentration of BANDs in the layer adjacent to the interface with ETL. Thus, for the MAPbI$_3$ with the conduction band effective mass of $m^* = 0.2m_0$[10] and the density of states $N_C = 2x10^{18}$ $cm^{-3}$ the CBE is closely pinned to the Fermi energy for the defect concentration $N_d > 10^{18}$ $cm^{-3}$.

Similarly, the above considerations apply to the charge transfer between perovskite layer and an HTL. In this case, holes from BANDs in the neutral acceptor configuration are transferred to higher located VBE of the HTL creating a depletion region in the perovskite layer next to the HTL where all BANDs become acceptors. Consequently, the layer adjacent to HTL acquires properties of a *p*-type material with the perovskite VBE pinned to the Fermi energy.

To evaluate the effect of charge transfer and the formation of the *n*-type and *p*-type layers more closely, we solve the Poisson's equation for field distribution at the interface of two *n*-type (*p*-type) layers with specified band offsets. The details of the calculations are discussed in the Methods Section. To account for the constraints (see Eq. 8) on the charge transfer, we limit the calculations to the distance from the interface where the electron (hole) concentration in perovskite CBE (VBE) is equal to the concentration of electrons (holes) transferred to the ETL (HTL). Results of the modeling are shown in Fig. 3. The charge transfer stabilizes BANDs in the donor (acceptor) configuration, pinning the Fermi energy close to the perovskite CBE (VBE) on the ETL (HTL) side. Therefore, the electron (hole) transfer forms an *n*-type (*p*-type) layer next to the ETL (HTL) with the intrinsic region in between and the device acquires all the characteristics of the *n-i-p* solar cell with the open circuit voltage (*V$_{oc}$*) determined by the perovskite band gap[37].

Previous attempts to explain the formation of an *n-i-p* structure in perovskite films assumed diffusion of defects; donors to the ETL side and acceptors to the HTL side of the device[38]. The attractiveness of this model was that conceptually it could account for other observed effects such as hysteresis and poling induced junction formation. The main drawback of the diffusion model is that it is not obvious what is the driving force for the defect separation and *n-i-p* junction formation. Also, it requires unrealistically fast, reversible, room temperature mass diffusion over significant distances in a polycrystalline material. **The BANDs model has all the advantages of the diffusion model without the need for any mass diffusion.** The formation of the *n*-type or *p*-type layer is associated with a *local* defect relaxation into donor-like or acceptor-like configuration driven by an excess of holes or electrons at the ETL or HTL side of the PV device.

The results in Fig. 3 indicate that the location of the perovskite CBE (VBE) relative to the Fermi energy does not significantly depend on the band offsets or the defect concentration as long as the defect concentration at the ETL (HTL) interface is larger than the perovskite conduction (valence) band density of states. These results explain why the $V_{oc}$ of perovskite PVs is largely independent of the location of the CBE of ETL and VBE of HTL. However, as is seen in Fig. 3 the shape i.e. the height and the width of the interface barrier depend on both band offsets and the total BAND concentration. Figure 3(c) shows the calculated barrier height ($\Phi_{ETL}$) and barrier width ($w_n$) as functions of the BANDs concentration. As expected, the height of the barrier is increasing with increasing band offsets whereas the width is increasing with increasing band offsets and decreasing BANDs concentration. This has important consequences for the solar cell performance as the series resistance for electrons transferring across the barrier is strongly dependent on the barrier height and the barrier width[39] which, as is seen in Fig. 3(d), depend on the concentration of BANDs indicating that any change in the net donor concentration will result in a significant change of the perovskite/ETL contact resistance. However, the effects will be less pronounced at large defect concentrations where the interface barrier resistance is small relative to other contributions to the total series resistance of the PVSC.

The discussed above mechanism of the *n-i-p* PVSC formation requires a sufficient electron (hole) transfer from perovskite to the ETL (HTL). This necessitates large enough concentration of BANDs in the perovskite absorber and sufficiently high density of unoccupied states in the ETL (HTL) to support a charge transfer large enough to pin the CBE (VBE) to the Fermi energy. Thus, if the concentration of BANDs close to the interface with ETL (HTL) is too low, the highly defected layer at the interface is too thin or the density of unoccupied states in the CBE (VBE) of ETL (HTL) is too low then the concentration of transferred electrons (holes) will not be high enough to pin the perovskite CBE (VBE) to the Fermi energy. This will result in a PVSC with lower $V_{oc}$, larger barrier and thus also larger series resistance. Another important conclusion that can be drawn from the results in Fig. 3 is that although the $V_{oc}$ is not sensitive to the location of the CBE (VBE) of ETL (HTL) the actual thickness and the height of the interfacial barrier and thus also the series resistance for the charge collection depends on the band offsets between charge transporting layers and the perovskite band edges. Also, it is important to note that the charge transfer can be affected by the presence of the interface charge, which can affect the shape of the interfacial barrier. It will be discussed in the following section of interface charge effect.

The dynamic nature of the formation of *n*-type (*p*-type) layers at ETL (HTL) interface makes the performance of perovskite solar cells sensitive to any process affecting the charge transfer between perovskite and ETL (HTL). A straightforward explanation for the origin of the hysteresis effects and the UV induced degradation of perovskite based solar cells will be explored in the flowing sections.

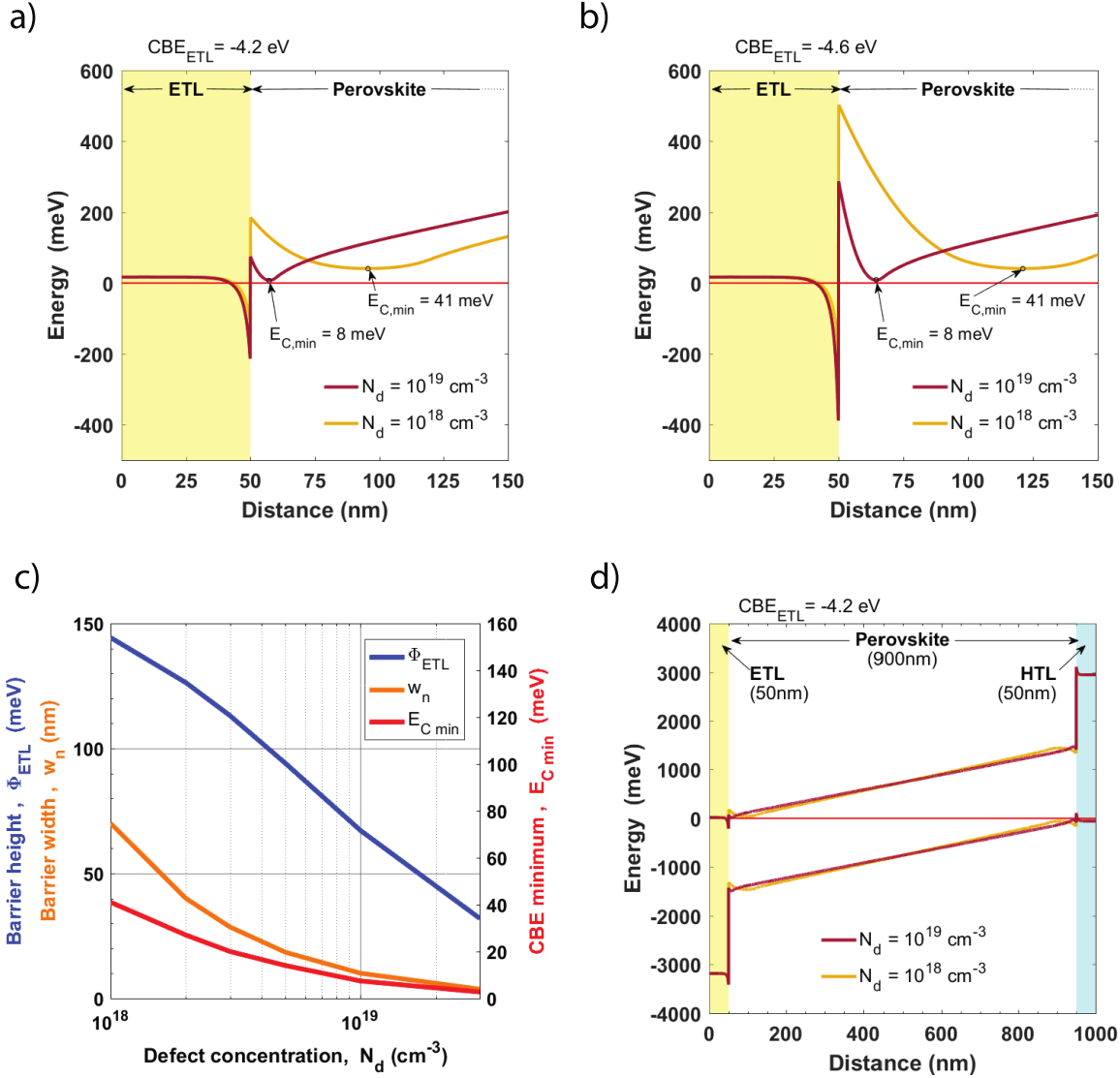

**Figure 3** Calculated dependence of the CBE energy for two different concentrations of BANDs and two different band offsets between perovskite and the ETL, 0.3 eV (**a**) and 0.7 eV (**b**). (**c**) Dependencies of the barrier height ($\Phi_{ETL}$), barrier thickness ($w_n$) and the energy separation between perovskite CBE and the Fermi energy ($E_{C\,min}$) on the BANDs concentration. (**d**) Band diagram of the complete perovskite solar cell structure, with an effective *n-i-p* configuration, n for two different concentrations of BANDs.

## I. Effects of the interface charge

In the BAND model for the perovskite PV formation, we have assumed that there is no additional charge at the interfaces of perovskite and charge transporting layers. However, as is well known, the presence of dangling bonds at the interfaces produces additional charge that pins the Fermi energy at the $E_{FS}$. Figure 4 shows the calculated band diagram for the case of perovskite/ETL interface with charge density required to pin the Fermi energy at $E_{FS}$ located at -4.9 eV below the vacuum level. It is seen that for the case of 0.3 eV band offset, adding the interface charge increases the barrier height (Fig. 4(a)). This, in turn, will increase the series resistance and have a detrimental effect on the PV performance. This effect can be mitigated by reducing the band offset. Thus, as is illustrated in Fig. 4(b), a very favorable band alignment with a very small barrier is realized for the zero-band offset between perovskite and ETL. In this case the formation of the *n*-type layer in the perovskite is associated with charge transfer to the depletion layer in the ETL.

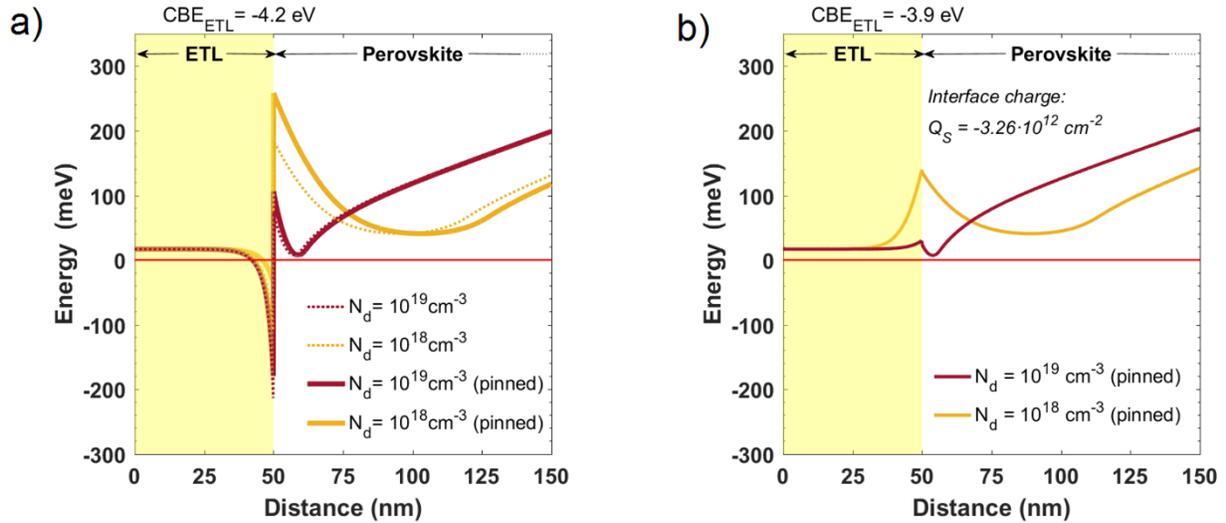

**Figure 4 Calculated band structure for the perovskite ETL interfaces with and without interface charge**. (**a**) For the 0.3 eV perovskite/ETL band offset case (i.e. $CBE_{ETL}$ = -4.2eV), adding the interface charge density sufficient to pin the Fermi energy at $E_{FS}$ (solid lines) increases the barrier at the interface compared with the case of no interface charge (dotted lines). (**b**) A greatly reduced barrier height is found in the case of zero CBE band offset (i.e. $CBE_{ETL}$ = -3.9eV) and the interface charge density sufficient to pin the Fermi energy at $E_{FS}$. In this case, an almost ideal contact between perovskite and ETL is formed for a BANDs density of $10^{19}$ cm$^{-3}$

## II. Origin of Hysteresis Effects in Perovskite Solar Cells

Hysteresis in the current voltage (*J-V*) characteristics is one of the salient and routinely observed features of perovskite-based PVs[40-42]. The simplest form of hysteresis manifests itself as a dependence of the shape of *J-V* curves on the voltage scanning direction and the scanning rates[43]. Thus, as has been shown in numerous papers, scanning of the external voltage from the short circuit current ($J_{sc}$) to open circuit voltage ($V_{oc}$) conditions (forward bias) yields a poorer performance than scanning from $V_{oc}$ to $J_{sc}$ (reverse bias)[38,44-46]. The most evident and always observed cause of this difference is a reduced fill factor for the forward scans compared with reverse scans. The detrimental effects of hysteresis on PVSC performance have led to numerous studies of this effect. It has been shown that the magnitude of the hysteresis observed in *J-V* characteristics depends on the scan rate, temperature and the perovskite/ETL interface design.

The reactions 5a to 5f (main text) describing formation of the *n*-type layer on the ETL side and equivalent reactions for the formation of a *p*-type layer on the HTL side correspond to a PV device illuminated under open circuit conditions. In order to understand the origin of the hysteresis, we note that, as is shown in Fig. 3(a) and (b), the formation of the *n*-type layer close to ETL results in a barrier and thus also electrical resistance for electron transfer from the perovskite absorber to the ETL. Under the open circuit condition, there is no current and thus there is no potential drop across the barrier. However, the barrier and the interface resistance make a difference when the device operates under the short circuit conditions with the maximum current $J_{sc}$ or under any conditions with a significant current density. Thus, as is schematically shown in Fig. 5(a) the $J_{sc}$ photo-current will result in a potential drop, $\Delta V_I$ across the interface barrier resistance. The change in the potential partially reverses the electron transfer between the perovskite and ETL by reversing the directions for the reactions 5.

$$e_{PVSK} \leftarrow e_{ETL} \tag{9a}$$

$$D^0 \leftarrow D^+ + e_{PVSK} \tag{9b}$$

$$A^0 \Leftarrow D^0 \tag{9c}$$

$$h_{PVSK} + A^- \leftarrow A^0 \tag{9d}$$

$$h\nu \leftarrow e_{PVSK} + h_{PVSK} \tag{9e}$$

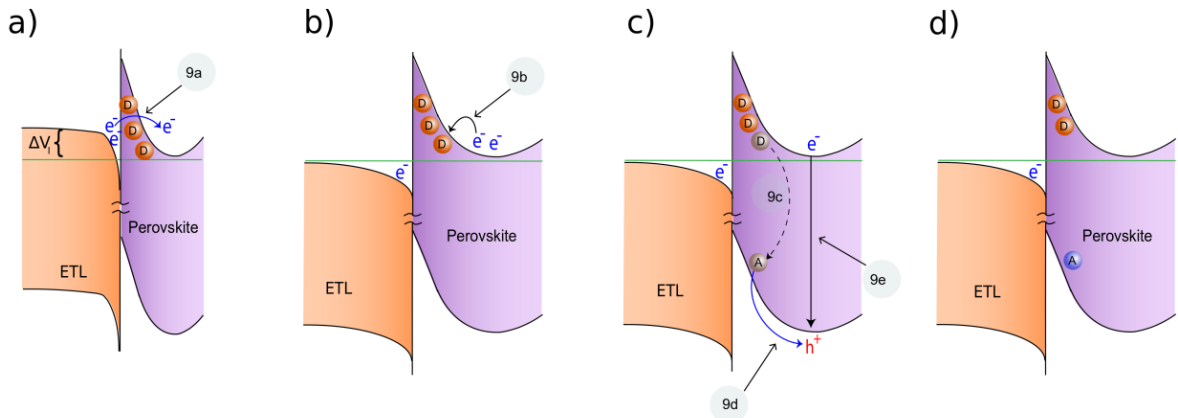

**Figure 5 Mechanism of the photocurrent induced reverse charge transfer resulting in an increase of the interfacial barrier width and series resistance**. **a)** Photocurrent induced potential drop transfers electrons from the ETL to perovskite (reaction 9a). **b)** The electron neutralizes a BAND donor (9b), **c)** The neutral donor transforms into neutral acceptor configuration (9c). Ionization of the acceptor produces a negatively charged acceptor and a hole in the valence band (9d). The hole recombines with an electron in the conduction band (9e). The process replaces

positively charged donor with a negatively charged acceptor reducing the net positive charge in the depletion region leading to a thicker, higher interface barrier and higher series resistance.

The above reverse reactions result in a new equilibrium with increased concentration of negatively charged BAND acceptors (reactions 9c and 9d) that compensate BAND donors reducing the effective concentration of the positive charge in the depletion layer adjacent to the ETL. However, as is evident from Fig. 3, a smaller net charged donor density in the depletion region leads to higher and thicker barrier for electron transport to the ETL and thus also higher series resistance.

The overall conclusion of this analysis is that the series resistance and thus also the shape of a perovskite PV *J-V* curves depend on the initial bias conditions for the *J-V* scan. Therefore, as is shown in Fig. 6, the increase of the barrier width and the series resistance under short circuit conditions will lower the fill factor and degrade the performance for the forward scan from $J_{sc}$ to $V_{oc}$. In contrast, illumination of the device under open voltage conditions transfers electrons back to the ETL and reduces the barrier. Consequently, the device will show improved performance with larger fill factor for the reverse scan from to $V_{oc}$ to $J_{sc}$. The discussed above mechanism provides a straightforward explanation for the effects of bias conditions during light soaking on the performance of a PVSC[47], where it has been clearly shown that a forward bias scan decreases and a reverse bias scan increases the fill factor (see Fig 7 of reference 14). Also, it explains why hysteresis effects depend on several factors such as *J-V* scanning rates, temperature, concentration of BANDs and band offsets. For example, as is shown in Fig. 1(b) the transformation between donor and acceptor BAND configuration requires thermal excitation over the configurational barrier meaning that the reaction represented by 9c and thus also the hysteresis effect are thermally activated and can be suppressed at low temperature. This observation is in excellent agreement with experimental data indicating that the hysteresis effects are rapidly decreasing with decreasing temperature[48].

The above discussion helps to understand and provides justification for the methods used to reduce or even eliminate the hysteresis in *J-V* characteristics. The lower fill factors for the forward *J-V* scans originate from an increase in the barrier height and thickness and thus also the series resistance for the scans starting from $J_{sc}$ bias conditions. This is especially critical when the concentration of defects is low and/or the defected layer is too thin and the corresponding band edge is only weakly pinned to the Fermi energy. In such case any back-charge transfer and reduced effective space charge in the depletion layer occurring during $J_{sc}$ bias will result in a significant change of the interface barrier for the charge collection and thus also larger series resistance.

Devising a structure that has a thick layer with a high concentration of BANDs will result in a strong pinning of the band edge to the Fermi energy making the contact resistance less sensitive to a reduction of effective BAND donor concentration at the interface. The above arguments provide an explanation for the experimentally observed reduction of the hysteresis in the PVSC device structures with mesoporous $TiO_2$ ETL compared with devices with planar $TiO_2$ ETL[44,47]. The mesoporous structure greatly enlarges the perovskite/$TiO_2$ contact area and thus also electron transfer from perovskite to TiO$_2$. Consequently, the whole thick mesoporous layer becomes *n*-type with the CBE strongly pinned to the Fermi energy and with a low interface resistance making the structure much less sensitive to the back-charge transfer under $J_{sc}$ bias.

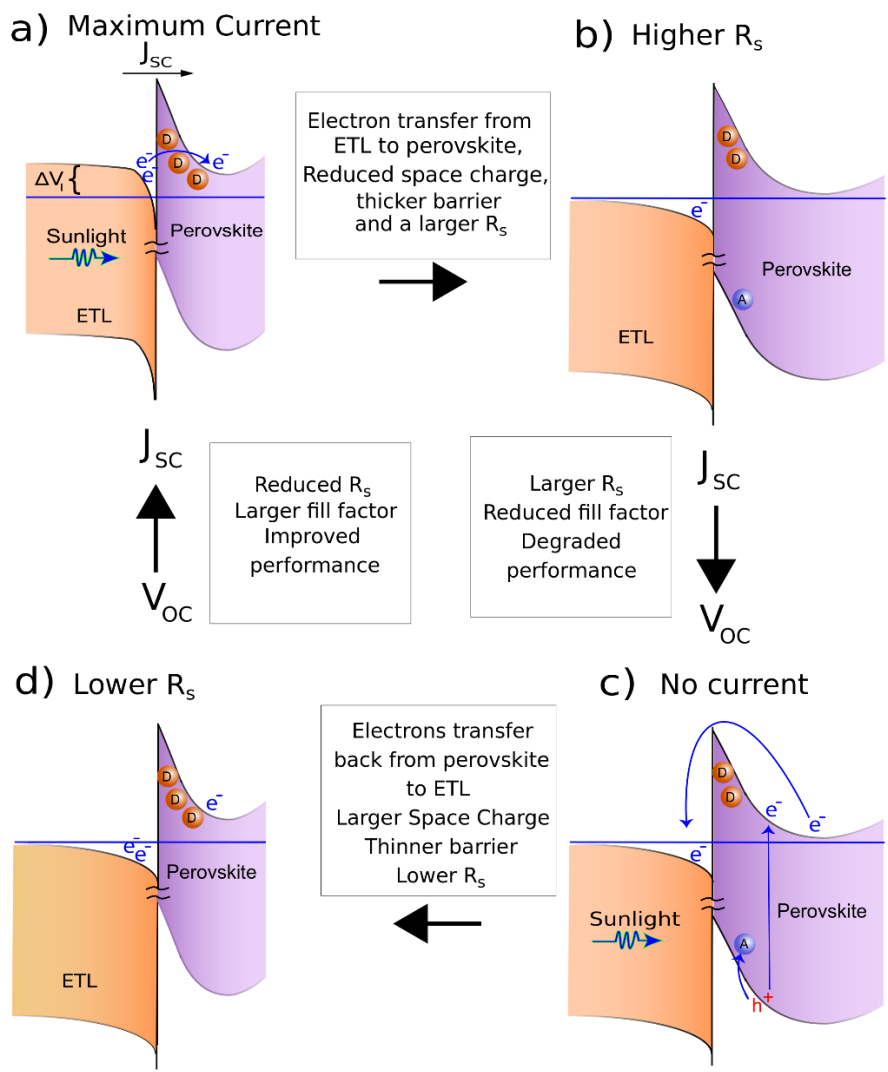

**Figure 6 Schematic representation of the hysteresis effects in a PVSC**. (**a**) Under nonzero current conditions the potential drop at the interface transfers electrons from the ETL to the perovskite layer resulting in (**b**) partial transformation of BANDs from donor to acceptor configuration results in lower depletion charge density, thicker barrier, larger series resistance and degraded *J-V* characteristic for $J_{sc}$ to $V_{oc}$ scan. (**c**) Illumination under open circuit (zero current) condition transfers electrons back to ETL restoring the original shape of the interface barrier shown in (**d**) leading to improved *J-V* characteristic for $V_{oc}$ to $J_{sc}$ scan.

### III. Ultraviolet Light Induced Perovskite Solar Cell Degradation

It has been found very early that an exposure to the UV radiation degrades properties of the PVSC structures with TiO$_2$ ETLs[49-51]. Originally this effect has been entirely attributed to photo-degradation of TiO$_2$[52]. However, in a most recent study, it was shown that under strictly controlled environmental conditions, the initial phase of the UV induced degradation of the PVSC performance is optically reversible through illumination with the solar spectrum photons[53]. It was also shown that a permanent, irreversible degradation associated with photoelectrochemical reactions in TiO$_2$ occurs for the UV exposure times longer than 200 hrs. Here we show that the BAND model provides a simple and elegant explanation of the initial, optically reversible UV degradation of perovskite solar cells.

The UV degradation process schematically shown in Fig. 7, is described by the sequence of reactions,

$$h\nu(UV) \rightarrow e_{ETL} + h_{ETL} \quad (10a)$$

$$e_{ETL} \rightarrow e_{PVSK} \quad (10b)$$

$$e_{ETL} + h_{ETL} \rightarrow h\nu^* \quad (10c)$$

$$e_{PVSK} + D^+ \rightarrow D^0 \quad (10d)$$

$$D^0 \Rightarrow A^0 \quad (10e)$$

$$A^0 \rightarrow A^- + h_{PVSK} \quad (10f)$$

$$e_{PVSK} + h_{PVSK} \rightarrow h\nu^* \quad (10g)$$

Note that the PVSC degradation process is again associated with a release (10g) of the electronic energy stored at the interface. The recombination of electrons and holes represented by reactions (10f) and (10g) can occur through a radiative or non-radiative channel. The final result of the reaction chain 10a to 10g is the transformation of one of the positively charged donors into a compensating negatively charged acceptor leading to lower net concentration of the positive charges in the perovskite depletion layer at the TiO$_2$/perovskite interface. This, in turn, leads to a thicker barrier and higher series resistance resulting in a degraded performance of the PVSC that primarily manifests itself in a reduction of the fill factor[53]. This is in excellent agreement with experiments showing that in an initial stage (up to 200 hrs) an UV illumination results in a pronounced increase of the series resistance and reduction of the fill factor but does not significantly affect the short circuit current and open circuit voltage.

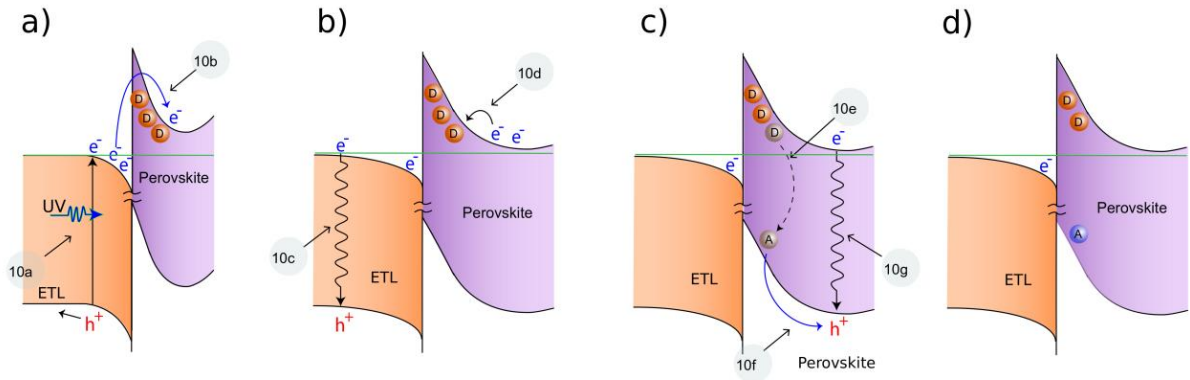

**Figure 7** UV induced degradation of a PVSC with TiO$_2$ ETL. (**a**) Illumination with UV light produces electron-hole pair in the *n*-type TiO$_2$ (reaction 10a). The photoexcited electron transfers to the perovskite (10b) and the hole diffuses away from the interface and recombines with an electron in n-type TiO$_2$ (10c). (**b**) The electron transferred to the perovskite neutralizes a BAND donor (10d) that transforms to the neutral acceptor configuration (10e). (**c**) Ionization of the acceptor produces a negatively charged acceptor and a free hole (10f). The hole recombines with an electron (10g). (**d**) This leads to reduced concentration of a positive charge in the depletion layer resulting in a thicker barrier, larger series resistance and smaller fill factor. The sunlight induced recovery follows the reactions 5a to 5f for the light induced *n*-type layer formation.

The sunlight induced recovery of a UV degraded cell is easy to understand based on the previous considerations of the light induced charge transfer and formation of the *n*-type layer next to the TiO$_2$ ETL under open circuit condition. According to the reactions 10a to 10f the original *n*-type layer with the Fermi energy strongly pinned to the perovskite CBE is restored when the sunlight excites electron hole pairs in the perovskite. It should be noted that illumination with the sunlight spectrum extending from UV to infrared excites electron hole pairs in both TiO$_2$ ETL and the perovskite. Therefore, under the solar spectrum illumination the cell performance is determined by a balance between degradation induced by UV photons absorbed in TiO$_2$ and recovery induced by lower energy photons absorbed in the perovskite. The UV induced degradation can be eliminated by utilizing ETL with a larger energy gap, outside the solar spectrum. This is supported by a recent work in which it was shown that replacing of TiO$_2$ with larger gap SnO$_2$ greatly improves the initial stability of perovskite PVs[54].

## Conclusion

We have developed a model of hybrid organic-inorganic perovskite solar cell operation. The model is based on a resonant interaction between free carriers and a system of bistable amphoteric defects. The nature, donor or acceptor, and the concentration of the defects is directly dependent on the Fermi energy and thus also charge carrier concentration. Charge transfer between the perovskite and electron and hole transporting layers affects the free charge distribution and produces a dynamic *n-i-p* junction in the perovskite absorber. The open circuit voltage, $V_{oc}$ of the device is determined by the perovskite band gap rather than the band offsets between charge transporting layers. The dynamic nature of the junction makes the perovskite solar cell susceptible to external perturbations. The model of the cell operation explains observed hysteresis of *J-V* characteristics. It also accounts for the reversible UV induced degradation of perovskite cells with a $TiO_2$ electron transport layer. The more general significance of our work is that we have identified a material system in which defects not only determine materials properties but also play a central, indispensable role in the device operation. In a forthcoming paper we will show that the presence of BANDs on surfaces and in the bulk have far reaching consequences for understanding optical and transport properties of the bulk perovskite materials, including poling effects, balanced charged transport and excitation intensity dependent photoluminescence.

**Acknowledgments**

The authors acknowledge funding support from the Singapore National Research Foundation (NRF) through the Singapore Berkeley Research Initiative for Sustainable Energy (SinBeRISE) Program" and from the National Research Foundation, Prime Minister's Office, Singapore under its Competitive Research Program (CRP Award No. NRF-CRP14-2014-03). G. Han acknowledges the funding support from Singapore Ministry of Education Tier 2 project MOE2016-T2-2-012. I. Rey-Stolle acknowledges UPM for the funding of his sabbatical stay at LBNL and also the Spanish Ministry of Economy and Competitiveness (MINECO) for funding through project TEC2015-66722-R. The authors acknowledge P.C. Harikesh for help with the illustrations.

**Notes**

The authors declare no competing financial interest.

**References:**

1   Kojima, A., Teshima, K., Shirai, Y. & Miyasaka, T. Organometal halide perovskites as visible-light sensitizers for photovoltaic cells. *J. Am. Chem. Soc.* **131**, 2, doi:10.1021/ja809598r (2009).
2   Kim, H. S. *et al.* Lead iodide perovskite sensitized all-solid-state submicron thin film mesoscopic solar cell with efficiency exceeding 9%. *Sci. Rep.* **2**, 591, doi:10.1038/srep00591 (2012).
3   Lee, M. M., Teuscher, J., Miyasaka, T., Murakami, T. N. & Snaith, H. J. Efficient hybrid solar cells based on meso-superstructured organometal halide perovskites. *Science* **338**, 643-647, doi:10.1126/science.1228604 (2012).

4	Yang, W. S. *et al.* Iodide management in formamidinium-lead-halide–based perovskite layers for efficient solar cells. *Science* **356**, 1376-1379, doi:10.1126/science.aan2301 (2017).
5	Sum, T. C. & Mathews, N. Advancements in Perovskite Solar Cells: Photophysics behind the Photovoltaics. *Energy & Environmental Science* **7**, 2518-2534, doi:10.1039/c4ee00673a (2014).
6	Snaith, H. J. Perovskites: The Emergence of a New Era for Low-Cost, High-Efficiency Solar Cells. *J. Phys. Chem. Lett.* **4**, 3623-3630, doi:10.1021/jz4020162 (2013).
7	Jung, H. S. & Park, N. G. Perovskite solar cells: from materials to devices. *Small* **11**, 10-25, doi:10.1002/smll.201402767 (2015).
8	Ball, J. M. & Petrozza, A. Defects in perovskite-halides and their effects in solar cells. *Nat. Energy* **1**, 16149, doi:10.1038/nenergy.2016.149 (2016).
9	Green, M. A., Ho-Baillie, A. & Snaith, H. J. The emergence of perovskite solar cells. *Nat. Photonics* **8**, 506-514, doi:10.1038/nphoton.2014.134 (2014).
10	Brenner, T. M., Egger, D. A., Kronik, L., Hodes, G. & Cahen, D. Hybrid organic—inorganic perovskites: low-cost semiconductors with intriguing charge-transport properties. *Nature Rev. Mater.* **1**, 15007, doi:10.1038/natrevmats.2015.7 (2016).
11	Yin, W. J., Shi, T. & Yan, Y. Unique Properties of Halide Perovskites as Possible Origins of the Superior Solar Cell Performance. *Adv. Mater.* **26**, 4653-4658, doi:10.1002/adma.201306281 (2014).
12	Agiorgousis, M. L., Sun, Y.-Y., Zeng, H. & Zhang, S. Strong Covalency-Induced Recombination Centers in Perovskite Solar Cell Material CH3NH3PbI3. *J. Am. Chem. Soc.* **136**, 14570-14575, doi:10.1021/ja5079305 (2014).
13	Du, M. H. Density Functional Calculations of Native Defects in CH3NH3PbI3: Effects of Spin–Orbit Coupling and Self-Interaction Error. *J. Phys. Chem. Lett.* **6**, 1461-1466, doi:10.1021/acs.jpclett.5b00199 (2015).
14	Brandt, R. E., Stevanović, V., Ginley, D. S. & Buonassisi, T. Identifying defect-tolerant semiconductors with high minority-carrier lifetimes: beyond hybrid lead halide perovskites. *MRS Commun.* **5**, 265-275, doi:10.1557/mrc.2015.26 (2015).
15	Walsh, A. & Zunger, A. Instilling defect tolerance in new compounds. *Nat. Mater.* **16**, 964, doi:10.1038/nmat4973 (2017).
16	Walukiewicz, W. Mechanism of Schottky barrier formation: The role of amphoteric native defects. *J. Vac. Sci. Technol. B* **5**, 1062-1067, doi:10.1116/1.583729 (1987).
17	Walukiewicz, W. Fermi level dependent native defect formation: Consequences for metal–semiconductor and semiconductor–semiconductor interfaces. *J. Vac. Sci. Technol. B* **6**, 1257-1262, doi:10.1116/1.584246 (1988).
18	Walukiewicz, W. Mechanism of Fermi-level stabilization in semiconductors. *Phys. Rev. B* **37**, 4760-4763, doi:10.1103/PhysRevB.37.4760 (1988).
19	Alberi, K. & Scarpulla, M. A. Suppression of compensating native defect formation during semiconductor processing via excess carriers. *Sci. Rep.* **6**, 27954, doi:10.1038/srep27954 (2016).
20	Walukiewicz, W. Intrinsic limitations to the doping of wide-gap semiconductors. *Physica B* **302**, 123-134, doi:https://doi.org/10.1016/S0921-4526(01)00417-3 (2001).
21	Baraff, G. A. & Schluter, M. Bistability and Metastability of the Gallium Vacancy in GaAs: The Actuator of *EL* 2? *Phys. Rev. Lett.* **55**, 2340-2343, doi:10.1103/PhysRevLett.55.2340 (1985).
22	Van de Walle, C. G. & Neugebauer, J. Universal alignment of hydrogen levels in semiconductors, insulators and solutions. *Nature* **423**, 626, doi:10.1038/nature01665 (2003).
23	Noel, N. K. *et al.* Enhanced Photoluminescence and Solar Cell Performance via Lewis Base Passivation of Organic–Inorganic Lead Halide Perovskites. *ACS Nano* **8**, 9815-9821, doi:10.1021/nn5036476 (2014).

24     Sadoughi, G. *et al.* Observation and Mediation of the Presence of Metallic Lead in Organic–Inorganic Perovskite Films. *ACS Appl. Mater. Interfaces* **7**, 13440-13444, doi:10.1021/acsami.5b02237 (2015).
25     Han, G. *et al.* Facile Method to Reduce Surface Defects and Trap Densities in Perovskite Photovoltaics. *ACS Appl. Mater. Interfaces* **9**, 21292-21297, doi:10.1021/acsami.7b05133 (2017).
26     Rong, Y., Liu, L., Mei, A., Li, X. & Han, H. Beyond Efficiency: the Challenge of Stability in Mesoscopic Perovskite Solar Cells. *Adv. Energy Mater.* **5**, 1501066, doi:10.1002/aenm.201501066 (2015).
27     Han, G. *et al.* Towards high efficiency thin film solar cells. *Prog. Mater Sci.* **87**, 246-291, doi:http://dx.doi.org/10.1016/j.pmatsci.2017.02.003 (2017).
28     Mahmood, K., Swain, B. S. & Amassian, A. 16.1% Efficient Hysteresis-Free Mesostructured Perovskite Solar Cells Based on Synergistically Improved ZnO Nanorod Arrays. *Adv. Energy Mater.* **5**, 1500568, doi:10.1002/aenm.201500568 (2015).
29     Correa Baena, J. P. *et al.* Highly Efficient Planar Perovskite Solar Cells through Band Alignment Engineering. *Energy & Environmental Science* **8**, 2928-2934, doi:10.1039/C5EE02608C (2015).
30     Xiong, J. *et al.* Efficient and non-hysteresis CH3NH3PbI3/PCBM planar heterojunction solar cells. *Org. Electron.* **24**, 106-112, doi:http://dx.doi.org/10.1016/j.orgel.2015.05.028 (2015).
31     Ryu, S. *et al.* Voltage output of efficient perovskite solar cells with high open-circuit voltage and fill factor. *Energy & Environmental Science* **7**, 2614-2618, doi:10.1039/C4EE00762J (2014).
32     Qin, P. *et al.* Inorganic hole conductor-based lead halide perovskite solar cells with 12.4% conversion efficiency. *Nat. Commun.* **5**, 3834, doi:10.1038/ncomms4834 (2014).
33     Arora, N. *et al.* Perovskite solar cells with CuSCN hole extraction layers yield stabilized efficiencies greater than 20%. *Science*, doi:10.1126/science.aam5655 (2017).
34     Christians, J. A., Fung, R. C. & Kamat, P. V. An inorganic hole conductor for organo-lead halide perovskite solar cells. Improved hole conductivity with copper iodide. *J. Am. Chem. Soc.* **136**, 758-764, doi:10.1021/ja411014k (2014).
35     Mei, A. *et al.* A hole-conductor-free, fully printable mesoscopic perovskite solar cell with high stability. *Science* **345**, 295-298, doi:10.1126/science.1254763 (2014).
36     Leijtens, T. *et al.* Carrier trapping and Recombination: the Role of Defect Physics in Enhancing the Open Circuit Voltage of Metal Halide Perovskite Solar Cells. *Energy & Environmental Science* **9**, 3472-3481, doi:10.1039/C6EE01729K (2016).
37     Edri, E. *et al.* Elucidating the Charge Carrier Separation and Working Mechanism of CH3NH3PbI3-xClx Perovskite Solar Cells. *Nat. Commun.* **5**, 3461, doi:doi:10.1038/ncomms4461 (2014).
38     Zhang, Y. *et al.* Charge selective contacts, mobile ions and anomalous hysteresis in organic-inorganic perovskite solar cells. *Mater. Horiz.* **2**, 315-322, doi:10.1039/C4MH00238E (2015).
39     Chang, C. Y., Fang, Y. K. & Sze, S. M. Specific contact resistance of metal-semiconductor barriers. *Solid-State Electron.* **14**, 541-550, doi:https://doi.org/10.1016/0038-1101(71)90129-8 (1971).
40     Zhao, C. *et al.* Revealing Underlying Processes Involved in Light Soaking Effects and Hysteresis Phenomena in Perovskite Solar Cells. *Adv. Energy Mater.* **5**, 1500279, doi:10.1002/aenm.201500279 (2015).
41     Zhang, H. *et al.* Dynamic interface charge governing the current-voltage hysteresis in perovskite solar cells. *Phys. Chem. Chem. Phys.* **17**, 9613-9618, doi:10.1039/C5CP00416K (2015).
42     Tress, W. *et al.* Understanding the rate-dependent J-V hysteresis, slow time component, and aging in CH3NH3PbI3 perovskite solar cells: the role of a compensated electric field. *Energy & Environmental Science* **8**, 995-1004, doi:10.1039/C4EE03664F (2015).
43     Ono, L. K., Raga, S. R., Wang, S., Kato, Y. & Qi, Y. Temperature-dependent hysteresis effects in perovskite-based solar cells. *J. Mater. Chem. A* **3**, 9074-9080, doi:10.1039/C4TA04969A (2015).

44  Wu, B. *et al.* Charge Accumulation and Hysteresis in Perovskite-Based Solar Cells: An Electro-Optical Analysis. *Adv. Energy Mater.*, 1500829, doi:10.1002/aenm.201500829 (2015).
45  Snaith, H. J. *et al.* Anomalous Hysteresis in Perovskite Solar Cells. *J. Phys. Chem. Lett.* **5**, 1511-1515, doi:10.1021/jz500113x (2014).
46  Kim, H.-S. & Park, N.-G. Parameters Affecting I–V Hysteresis of CH3NH3PbI3 Perovskite Solar Cells: Effects of Perovskite Crystal Size and Mesoporous TiO2 Layer. *J. Phys. Chem. Lett.* **5**, 2927-2934, doi:10.1021/jz501392m (2014).
47  Unger, E. L. *et al.* Hysteresis and transient behavior in current-voltage measurements of hybrid-perovskite absorber solar cells. *Energy & Environmental Science* **7**, 3690-3698, doi:10.1039/C4EE02465F (2014).
48  Bruno, A. *et al.* Temperature and Electrical Poling Effects on Ionic Motion in MAPbI3 Photovoltaic Cells. *Adv. Energy Mater.* **7**, 1700265, doi:10.1002/aenm.201700265 (2017).
49  Li, W. *et al.* Enhanced UV-light stability of planar heterojunction perovskite solar cells with caesium bromide interface modification. *Energy & Environmental Science* **9**, 490-498, doi:10.1039/C5EE03522H (2016).
50  Guarnera, S. *et al.* Improving the Long-Term Stability of Perovskite Solar Cells with a Porous Al2O3 Buffer Layer. *J. Phys. Chem. Lett.* **6**, 432-437, doi:10.1021/jz502703p (2015).
51  Ito, S., Tanaka, S., Manabe, K. & Nishino, H. Effects of Surface Blocking Layer of Sb2S3on Nanocrystalline TiO2for CH3NH3PbI3Perovskite Solar Cells. *J. Phys. Chem. C* **118**, 16995-17000, doi:10.1021/jp500449z (2014).
52  Leijtens, T. *et al.* Overcoming ultraviolet light instability of sensitized TiO(2) with meso-superstructured organometal tri-halide perovskite solar cells. *Nat Commun* **4**, 2885, doi:10.1038/ncomms3885 (2013).
53  Lee, S.-W. *et al.* UV Degradation and Recovery of Perovskite Solar Cells. *Sci. Rep.* **6**, 38150, doi:10.1038/srep38150 (2016).
54  Christians, J. A. *et al.* Tailored interfaces of unencapsulated perovskite solar cells for >1,000 hour operational stability. *Nat. Energy* **3**, 68-74, doi:10.1038/s41560-017-0067-y (2018).
55  Tan, I. H., Snider, G. L., Chang, L. D. & Hu, E. L. A self‐consistent solution of Schrödinger–Poisson equations using a nonuniform mesh. *J. Appl. Phys.* **68**, 4071-4076, doi:10.1063/1.346245 (1990).